\documentclass[accepted, specialissue]{melba}

\usepackage{mwe} 

\usepackage{amsmath,amsfonts}
\usepackage[nolist]{acronym}
\usepackage{array}
\usepackage{graphicx}
\usepackage{booktabs}
\usepackage{stfloats}
\usepackage{float}
\newcolumntype{P}[1]{>{\centering\arraybackslash}p{#1}}
\newcolumntype{L}[1]{>{\raggedright\arraybackslash}p{#1}}

\acrodef{cdss}[CDSS]{clinical decision support system}
\acrodef{dss}[DSS]{decision support system}
\acrodef{hci}[HCI]{human-computer interaction}
\acrodef{jas}[JAS]{judge advisor system}
\acrodef{tcp}[TCP]{tumor cell percentage}
\acrodef{woa}[WoA]{weight of advice}

\melbaid{2026:007}  
\doi{10.59275/j.melba.2026-87b1}
\melbaauthors{Rosbach et al}  
\email{emely.rosbach@thi.de}
\volume{2026}
\firstpageno{126}  
\melbayear{2026}  
\datesubmitted{2025-07-15}  
\datepublished{2026-03-13}  

\melbaspecialissue{MELBA–BVM 2025 Special Issue}
\melbaspecialissueeditors{Andreas Maier, Thomas Deserno, Heinz Handels, Klaus Maier-Hein, Christoph Palm, Thomas Tolxdorff, Katharina Breininger}

\ShortHeadings{Stuck on Suggestions: Automation Bias and the Anchoring Effect in Computational Pathology}{Rosbach et al.}

\title{Stuck on Suggestions: Automation Bias, the Anchoring Effect, and the Factors That Shape Them in Computational Pathology}

\author{
	\firstname Emely \surname Rosbach\aff{1}\orcid{0009-0001-7526-9923},
	\name Jonas Ammeling\aff{1}\orcid{0000-0002-0335-1194}, \name Jonathan Ganz\aff{1}\orcid{0009-0008-1299-8716}, \name Christof Albert Bertram\aff{2}\orcid{0000-0002-2402-9997}, \name Thomas Conrad\aff{3}\orcid{0009-0007-5593-1522}, \name Andreas Riener \aff{1}\orcid{0000-0002-9174-8895}, \name Marc Aubreville\aff{4}\orcid{0000-0002-5294-5247}
}
\affiliations{
	\num 1 \addr Technische Hochschule Ingolstadt, Ingolstadt, Germany\\
	 \num 2 \addr University of Veterinary Medicine Vienna, Vienna, Austria \\
    \num 3 \addr Freie Universität Berlin, Berlin, Germany \\
	\num 4 \addr Flensburg University of Applied Sciences, Flensburg, Germany \\    
}

\abstract{
Artificial intelligence (AI)-driven clinical decision support systems (CDSS) hold promise to improve diagnostic accuracy and efficiency in computational pathology. However, collaboration between human experts and AI may give rise to cognitive biases, such as automation and anchoring bias, wherein users may be inclined to blindly adopt system recommendations or be disproportionately influenced by the presence of AI predictions, even when they are inaccurate. These biases may be exacerbated under time pressure, pervasive in routine pathology diagnostics, or shaped by individual user characteristics. To investigate these effects, we conducted a web-based experiment in which trained pathology experts (n = 28) estimated tumor cell percentages twice: once independently and once with the aid of an AI. A subset of the estimates in each condition was performed under time constraints. Our findings indicate that AI integration generally enhances diagnostic performance. However, it also introduced a 7\% automation bias rate, quantified as the number of accepted negative consultations, where a previously correct independent assessment gets overturned by inaccurate AI guidance. While time pressure did not increase the frequency of automation bias occurrence, it appeared to intensify its severity, as evidenced by a performance decline linked to increased automation reliance under cognitive load. A linear mixed-effects model (LMM) analysis, simulating weighted averaging, revealed a statistically significant positive coefficient for AI advice, indicating a moderate degree of anchoring on system output. This effect was further intensified under time pressure, suggesting that anchoring bias may become more pronounced when cognitive resources are limited. A secondary LMM evaluation assessing automation reliance, used as a proxy for both automation and anchoring bias, demonstrated that professional experience and self-efficacy were associated with reduced dependence on system support, whereas higher confidence during AI-assisted decision-making was linked to increased automation reliance. Together, these findings underscore the dual nature of AI integration in clinical workflows, offering performance benefits while also introducing risks of cognitive bias–driven diagnostic errors. As an initial investigation focused on a single medical specialty and diagnostic task, this study aims to lay the groundwork for future research to explore these phenomena across diverse clinical contexts, ultimately supporting the establishment of appropriate reliance on automated systems and the safe, effective integration of human–AI collaboration in medical decision-making.}


\begin{document}

\twocolumn[\maketitle]

\section{Introduction}
\enluminure{R}{ecent} advances in artificial intelligence (AI) research promise to transform the healthcare landscape by addressing pervasive clinical challenges such as staff shortages and rising workloads~\citep{Metter2019}. Enhancing both diagnostic accuracy and workflow efficiency~\citep{Pantanowitz2010}, AI ultimately holds potential to better patient outcomes. This proves particularly relevant in disciplines traditionally demanding for human practitioners, such as quantitative analysis of microscopic images, including biomarker scoring, identification of mitotic figures, or tumor cell percentage assessments, where deep learning approaches have demonstrated performance on par with, or even exceeding, that of medical experts~\citep{Aubreville2020, Aubreville2023, Bertram2022}. These developments have led to the emergence of new fields, like computational pathology, which aims to complement conventional microscopy in diagnostic practice.

However, the deployment of fully autonomous AI systems in clinical practice is currently constrained by the safety-critical nature of the healthcare domain and the unresolved legal complexities surrounding liability for machine-induced errors~\citep{Schemmer2022}. Instead, in line with the concept of hybrid intelligence, where human-AI collaboration yields outcomes surpassing the capabilities of either party individually~\citep{Dellermann2019}, AI is more appropriately utilized as a \ac{dss}. This approach leverages the advantages of AI integration, such as improved accuracy and efficiency, while ensuring that ultimate responsibility for diagnoses and treatment decisions remains with the medical professional. However, the necessity of keeping the human "in the loop" can introduce unforeseen challenges, as the sheer presence of AI-generated recommendations may influence medical professional's judgment and potentially introduce or exacerbate cognitive biases. Much like artificial intelligence, human intelligence revolves around solving interpolation problems, constructing generalized understanding from finite and often imperfect information. This process is inherently ill-posed and requires a form of regularization to yield meaningful conclusions, which can be understood as pattern recognition. While classical regularization can be made explicit, for example by minimizing the bending energy of a function, its mechanisms remain largely opaque in both human intelligence and modern AI approaches such as deep learning. Put simply, in humans, cognitive biases are systematic errors in thinking that arise when mental shortcuts, established patterns and heuristics, are relied upon to expedite decision-making. Although often useful, these shortcuts can lead to distorted perception, faulty reasoning, or illogical interpretation, potentially resulting in medical malpractice~\citep{Tversky1974}.

While most studies on \acp{cdss} emphasize improvements in overall performance~\citep{Goddard2012}, \ac{hci} research consistently highlights that human cognitive limitations can hinder users from fully capitalizing on these benefits~\citep{Chromik2021,rosbachCB,Fogliato2022,Solomon2014}. One of the most notable cognitive fallacies is automation bias, which describes the tendency to regard automated suggestions as infallible, with blind trust often replacing vigilant information seeking~\citep{Parasuraman2010}. This might lead to errors when issues go unnoticed because the system fails to flag them (omission errors) or when incorrect automation output is adopted without critical evaluation (commission errors)~\citep{Skitka2000}. A related phenomenon is the anchoring bias. Anchoring occurs when individuals assign disproportionate weight to the first piece of information they receive~\citep{Furnham2011}, such as machine-generated advice, which can influence and carry over into subsequent decision-making. These cognitive biases pose a significant challenge to the effectiveness of human–machine collaboration, as overreliance on automated cues may lead users to align with system predictions irrespective of their correctness~\citep{Spatola2024}. In the context of AI-assisted medical decision-making, this inability to disregard erroneous AI recommendations could potentially compromise the quality of patient care. Moreover, the manifestation of anchoring and automation bias can be further impacted by environmental factors such as time pressure, ubiquitous in routine pathology, as well as individual user traits, including professional experience or decision confidence~\citep{Furnham2011,Goddard2012}.

Empirical studies that place medical experts, the actual end users of~\acp{cdss}, at the center of investigations into automation and anchoring bias in human-AI collaboration remain scarce~\citep{Goddard2012, Lyell2017}. Notably, none have been conducted in the field of pathology, despite the anticipated widespread integration of AI in this discipline~\citep{song2023artificial}. While related work exists, important distinctions must be made. Our prior work~\citep{rosbachCB}, situated in computational pathology, examined confirmation bias in AI-assisted medical decision-making. In contrast, the present manuscript centers around automation and anchoring bias, targeting a distinct set of cognitive mechanisms characterized by undue reliance on algorithmic suggestions rather than selective information processing. Moreover, although the systematic review by Goddard et al.~\citep{Goddard2012} synthesizes several studies, including among others five~\citep{Berner2003,Bogun2005,Friedman1999,Mckibbon2006,Westbrook2005} that investigate automation bias in the form of negative consultations across various healthcare specialties, none of these are situated in the pathology domain. Instead, they primarily examine discrete decision-making scenarios in largely non–image-based diagnostic tasks and do not consider how contextual or individual factors shape automation or anchoring bias. Consequently, the manifestation and contextual modulation of these two cognitive fallacies in AI-aided pathology has yet to be examined.

To address this gap, a within-subject, four-condition online experiment was conducted with trained pathology experts to assess how the presence of AI support could induce automation and anchoring bias during decision-making in routine pathology diagnostics, while also evaluating how time pressure, professional experience, and decision confidence shape these effects. As part of this investigation, participating medical experts were tasked with estimating the \ac{tcp} on 20 image patches derived from various hematoxylin and eosin (H\&E)-stained tissue slides, sourced from multiple datasets. Initially, the task was performed independently to establish a baseline and, following a two-week wash-out period, was then repeated with the aid of an AI algorithm providing \ac{tcp} predictions. To simulate time pressure, half of the patches in each segment were evaluated under time constraints. This paper is an extension of ~\cite{RosbachAB}, which yielded the following key findings:
    \begin{itemize}
    \item Integration of AI support led to a statistically significant increase in overall performance, attenuating the adverse effects of time pressure observed in the baseline condition.
    \item Automation bias, defined as the acceptance of negative consultations, where an initially correct human assessment gets overturned by faulty AI guidance, was recorded with a frequency of approximately 7\%. This rate aligns with related research reporting negative consultation acceptance rates ranging from 6\% to 11\% across various medical disciplines~\citep{Goddard2012}.
    \item Although the occurrence frequency of automation bias did not increase between AI-aided evaluations with and without time constraints, its severity intensified under time pressure, evidenced by a greater deviation from ground truth values. Automation reliance emerged as a key factor in this effect. In general, dependence on AI advice was found to be statistically significantly higher when time was limited, suggesting that participants leaned on algorithmic support to alleviate the cognitive load induced by time stress, in turn amplifying the severity of automation bias.
\end{itemize}
In this work, we extend the analysis of the experimental data with a specific focus on anchoring bias, a cognitive phenomenon also closely associated with undue automation reliance. This investigation revealed that:
\begin{itemize}
    \item The presence of AI recommendations induces a measurable anchoring effect. A linear mixed-effects model (LMM) simulating weighted averaging showed a statistically significant positive coefficient of moderate magnitude for the AI prediction, indicating it exerted nearly equal influence as the initial human estimate in shaping the final \ac{tcp} assessment, thus confirming the presence of anchoring bias.
    \item Time stress was found to intensify the observed anchoring effect, as reflected by the greater model coefficient for the AI prediction under time constraints. Despite this, participants’ independent evaluations remained the slightly more influential factor across both time pressure conditions.
    \item As for contextual influences, automation reliance, indicative of both automation and anchoring bias, increased under time pressure as well as with higher decision confidence during AI-assisted evaluations. Conversely, greater professional experience and decision confidence during baseline assessments, reflective of self-assuredness, appeared to mitigate dependence on AI input.
\end{itemize}
To summarize, the contributions of this research are as follows:
\begin{enumerate}
    \item We demonstrate the presence of automation bias and the anchoring effect in AI-assisted medical decision-making within computational pathology, primarily triggered through undue reliance on algorithmic suggestions. To the best of our knowledge, this is the first study to systematically investigate these biases in this context. Our findings highlight potential pitfalls that must be addressed prior to widespread deployment of AI-based \ac{cdss} in diagnostic practice.
    \item We contribute to the limited number of human-AI interaction studies that involve pathologists, the future end users of AI tools for pathology diagnostics, by engaging them in a routine task. This approach enhances the external validity and practical relevance of our findings.
    \item To our knowledge, we are the first to investigate how several contextual factors, including time pressure, professional experience, and decision confidence, affect the manifestation of automation and anchoring bias during human-AI collaboration in pathology. Our results indicate that automation reliance, a construct associated with both automation and anchoring bias, increases under time pressure and with elevated decision confidence during AI-assisted evaluations, while professional experience and self-assuredness seem to mitigate this dependence. These insights contribute to a deeper understanding of cognitive biases and their underlying mechanisms in AI-supported medical decision-making.
\end{enumerate}
    
\section{Background and Related Work}
\subsection{Automation Bias}
The concept of automation bias was first introduced by~\cite{Mosier2018}, defining it as the tendency to substitute effortful information processing with the heuristic use of automated cues, resulting in overreliance on automation. Automation bias typically manifests in form of omission errors, which arise when issues go unnoticed due to the system's failure to flag them, or commission errors occurring when users uncritically accept incorrect system output~\citep{Skitka2000}.

\paragraph{Measuring Automation Bias}
In their review,~\cite{Goddard2012} propose quantifying automation bias via the rate of commission errors, specifically through the acceptance of negative consultations, instances in which an initially accurate human judgment is overridden by erroneous system output. This approach enables a more precise and controlled measurement of automation bias by isolating it from other confounding cognitive fallacies. Originating in the aviation domain, automation bias has been observed across a wide range of \ac{hci} contexts, yet it has received comparatively limited empirical investigation within healthcare settings~\citep{Goddard2014}. Meta-analysis of this limited body of research, encompassing specialties such as electrocardiogram interpretation and drug administration, revealed that erroneous \acp{cdss} increase the likelihood of an incorrect clinical judgment by 26\%~\citep{Lyell2017}.

\paragraph{Impact of Situational and Individual Factors}
Various potential mediators of automation bias have been posited, e.g., task inexperience, overconfidence, cognitive load, system trust, and interface design~\citep{Goddard2012}. However, literature remains inconclusive whether (and in what context) these factors consistently contribute to automation bias~\citep{Lyell2017EP}. While prevailing human factors research primarily links automation bias occurrence to multitasking environments, it has also been observed in single-task settings, suggesting that task-specific features, such as the complexity of verifying automated output, may themselves be contributing factors~\citep{Lyell2017EP}. In healthcare, most existing empirical investigations of automation bias are centered around discrete decision-making~\citep{Kucking2024,Bogun2005,Friedman1999,Mckibbon2006}, such as binary judgments about whether to prescribe a particular medication, leaving continuous decisions comparatively underexplored. Similar ambiguity surrounds the role of time constraints in the manifestation of automation bias. Theoretical frameworks propose that time pressure increases cognitive load, thereby fostering greater reliance on automated systems~\citep{Goddard2012}. However, findings in related \ac{hci} research remain mixed: some studies report increased automation dependence under time constraints~\citep{Rice2008}, while others observe reduced reliance in time-critical scenarios~\citep{Rieger2022}. This discrepancy suggests that the influence of time pressure on automation bias may be task- or context-dependent.

Turning to individual characteristics, the theory of technology dominance (TTD)~\citep{Arnold1998} suggests that users with limited task experience are more prone to overrely on automation. This notion is supported by empirical findings e.g.,~\cite{Dreiseitl2005} observed that less experienced physicians were more likely to revise their initial diagnoses in response to conflicting suggestions from decision support systems. In the same vein, task-related confidence often referred to as self-efficacy can also shape automation dependence, as less confident individuals tend to seek greater amounts of advice~\citep{Gino2012}. Substantiating this point,~\cite{Lee1992} identified a trade-off between trust in automation and self-confidence: when users place greater trust in automation than in their own judgment, they are more likely to align with system output, increasing the risk of overreliance. Conversely, users’ confidence in their own decisions tends to increase when ample information is available from multiple sources~\citep{Budescu2000}.

\subsection{The Anchoring Effect}\label{sec:bgAE}
In their pioneering work on judgment under uncertainty,~\cite{Tversky1974} first introduced the anchoring-and-adjustment heuristic, demonstrating that exposure to an irrelevant anchor biased subsequent judgments in its direction. More broadly, anchoring bias refers to the tendency for an initial piece of information, regardless of its accuracy or relevance, to disproportionately influence human decision-making~\citep{Furnham2011}. In \ac{hci} anchoring often manifests in the form of automation reliance, where users tend to fixate on external anchors, i.e., machine-generated advice even when they have the expertise or contextual information to independently assess a situation~\citep{Spatola2024}. Consequently, the anchoring effect is regarded as a facet of automation bias~\citep{Rastogi2022}, particularly when users fail to appropriately adjust away from faulty automation cues or reverse their initially correct judgments after exposure to erroneous system output (attitude change).

\paragraph{Measuring the Anchoring Effect}
The anchoring effect can be operationalized using methods adapted from advice-taking paradigms, such as the \ac{jas}~\citep{Sniezek1995}, which examines how individuals (judges) revise their initial assessments after receiving input from an external source (advisor). The \ac{woa}, as introduced by~\cite{Harvey1997} (see Equation~\ref{eq:woa}), is among the most widely used metrics for quantifying advice weighting within the \ac{jas} framework, bearing conceptual resemblance to the anchoring indices proposed by~\cite{Jacowitz1995}~\citep{Rebholz2024}. It measures the relative amount of judgmental shift $\omega_\textrm{A,ij}$ from the initial estimate~$I_\textrm{ij}$ to the final judgment~$F_\textrm{ij}$ of participant~$i=1,...,N$ for stimulus~$j=1,...,M$ that can be attributed to a single piece of new information or advice~$A_\textrm{ij}$.
\begin{equation}
\omega_\textrm{A,ij}=\frac{F_\textrm{ij}-I_\textrm{ij}}{A_\textrm{ij}-I_\textrm{ij}} 
\label{eq:woa}
\end{equation}
Thus, a \ac{woa} value of 0 indicates no adjustment toward the advice, whereas a value of 1 reflects complete adoption, with everything in between corresponding to partial weighting of the input. The \ac{woa} can also take on negative values, indicating an adverse reaction to external input, i.e., correcting ones own initial evaluation in the opposite direction to the tendency of the advice, indicating distrust in the recommendation, e.g., due to perceived bias. If the advice is well intentioned and independent from one's initial estimate, incorporating 50\% of the recommendation would generally represent rational use of advice~\citep{Bailey2023}. However, empirical research demonstrates that people typically adjust only about 30\% toward external guidance on average~\citep{Soll2009}.
In healthcare, a study investigating AI-assisted cancer risk assessments by general practitioners reported a mean \ac{woa} of approximately 0.54~\citep{Palfi2022}, suggesting a moderate anchoring effect on system predictions. Although the \ac{woa} provides insight into both the direction and magnitude of the judgmental adjustment, it presents limitations. Values below 0 or above 1, suggesting overadjustment away from or beyond the advice, are often winsorized in related literature to fit within the 0–1 range~\citep{Palfi2022, Logg2019, Rebholz2024}. While the majority of observed judgment shifts typically fall within the 0-1 interval~\citep{Harvey1997} and outlier handling through exclusion/clipping may facilitate parametric testing, this approach may skew the mean toward the center, potentially masking the true extent of the anchoring effect. Additionally, the \ac{woa} metric might yield undefined values due to zero division, such as when the initial estimate equals the advice.

Normalization of the \ac{woa} ($Rel_\textrm{AI,ij}$), as illustrated in the work of~\cite{Lee2022} (see Equation \ref{eq:relAI}), ensures that results are strictly confined between 0 (complete disregard of the external advice) and 1 (full adoption of the recommendation), thus providing a solution to the previously described problem of out-of-range values. However, since their proposed metric is based on absolute differences, it quantifies solely the magnitude of the response towards external input failing to retain information about the direction of change or to detect instances of overadjustment, thereby potentially producing ambiguous values.
\begin{equation}
Rel_\textrm{AI,ij} = \frac{\left | F_\textrm{ij}  - I_\textrm{ij} \right |}{\left | F_\textrm{ij}  - A_\textrm{ij}\right | + \left | F_\textrm{ij} - I_\textrm{ij} \right |} 
\label{eq:relAI}
\end{equation}
Weighted averaging, frequently used as a modeling technique for decision-making under uncertainty~\citep{Ahn2014}, presents an alternative approach. By solving Equation~\ref{eq:woa} for $F_\textrm{ij}$, the final judgment can be calculated as a combination of the initial estimate and the provided recommendation, each with its own, complementary, weight (see Equation~\ref{eq:weighted_avg}). This technique is commonly employed in related research in conjunction with linear regression analysis, with model coefficients reflecting weighting~\citep{Rebholz2024,Logg2019,Liberman2012}.
\begin{equation}
F_\textrm{ij} = (1-\omega_\textrm{A,ij})I_\textrm{ij}  + \omega_\textrm{A,ij}A_\textrm{ij}
\label{eq:weighted_avg}
\end{equation}
It should be noted that although regression-based approaches are generally preferable, they also have limitations, particularly when the judge and advisor are not independent, for example when both parties have access to the same information~\citep{Bonaccio2006}. In such cases, multicollinearity may arise if regression coefficients are used to estimate the relative contributions of these predictors to the final assessment, complicating the interpretability of the regression results.

\paragraph{Influences of Context and Individual Traits}
As a facet of automation bias, anchoring bias similarly demonstrates context-dependent responses to both environmental conditions and individual user traits. While theoretical frameworks propose that time pressure depletes cognitive resources, thereby increasing reliance on external cues such as anchors~\citep{Epley2006}, empirical findings are mixed. For example, a study by~\cite{Rastogi2022} on AI-assisted prediction of student performance found that time constraints reduced participants’ ability to adjust away from initial anchors, amplifying the anchoring effect. In contrast, research on consumer price judgments found no significant influence of time pressure on anchoring behavior, highlighting the variability of outcomes across task domains~\citep{Zong2022}.

Regarding user characteristics, the aforementioned theory of technology dominance posits a decrease in automation reliance with increasing professional experience~\citep{Arnold1998}. Thus it stands to reason that anchoring bias might follow a similar pattern. Corroborating this point, research by \cite{Rastogi2022} on AI-supported student performance evaluation highlights that participants with relevant task-specific knowledge were able to de-anchor more effectively from system predictions.

Self-efficacy, here used synonymously with confidence, also serves as a measure of subjective knowledge~\citep{Brucks1985}. As such, lower confidence in one's judgment is theorized to similarly increase an individuals’ susceptibility to the influence of external information~\citep{Lee2022}, while overconfidence may contribute to advice discounting~\citep{Palfi2022}. For instance,~\cite{Clavelle1980} report that professionals with higher decision confidence revised their judgments less frequently. In support of this, a study by~\cite{Wu2012} on online shopping behavior revealed that only consumers with low self-efficacy were susceptible to anchoring bias. Shifting the focus from pre- to post-anchor confidence, related research suggests that exposure to anchors systematically increases individuals' confidence in their final judgments~\citep{Smith2017}. In \ac{hci}, a study in the domain of visual analytics observed higher decision confidence ratings with automated support compared to the control group~\citep{Wesslen2018}.
 
\subsection{Healthcare: a Unique Domain for Decision-making}
Healthcare is a high-stakes environment characterized by direct responsibility for patient outcomes, ubiquitous time pressure, incomplete or conflicting information, and significant intra-/inter-subject variability among both patients and clinicians~\citep{Rundo2020}. In light of these contextual challenges, practitioners often report on balancing empirical evidence with their professional intuition~\citep{Hall2002}. According to the framework of Naturalistic Decision Making, intuition shaped through years of experience serves as a valuable tool for pattern recognition~\citep{Kahneman2009}. For instance, research on medical image annotation indicates that the sheer volume of visual data to be evaluated often compels annotators to rely on intuitive judgments, as they offer a rapid alternative to more deliberate, analytical reasoning~\citep{Leiser2023}. However, renowned scholars such as~\cite{Kahneman2009} as well as~\cite{Meehl1973} warn that practitioners' reliance on gut feelings may be problematic, emphasizing that even expert intuition is frequently shaped by heuristics and cognitive biases. This proves particularly relevant in the subdomain of pathology, where many diagnostic processes rely heavily on the interpretation of often complex visual data, such as during tissue morphology assessment. The inherent difficulty humans face with precise visual quantification can manifest as high interobserver variability, partially driven by cognitive biases that emerge when individuals rely on intuition to reduce cognitive strain in demanding tasks~\citep{Aeffner2017, Viray2013}. Echoing these concerns, cognitive biases are increasingly being recognized as significant contributors to diagnostic errors. A related study found that cognitive factors, including cognitive biases, were implicated in up to 92\% of self-reported medical errors in emergency medicine~\citep{Okafor2016}. What distinguishes healthcare from other expert domains is that errors often arise not from a lack of knowledge, but from a flawed approach in judgment and decision-making~\citep{Graber2005}.

Although the concept of cognitive biases has been established for nearly 50 years~\citep{Tversky1974}, research examining these cognitive fallacies during interactions with (AI-based) \acp{dss} has only gained prominence with the recent surge in AI development. This increased relevance stems not only from advances in AI capabilities, but also from the critical role of system design: many AI systems, such as chatbots, exhibit anthropomorphic features in appearance or communication style, which can elicit inappropriate trust and subtly influence user judgment. Furthermore, because AI reasoning differs fundamentally from human cognitive processes, AI-generated advice is often perceived and engaged with differently, especially when distinctive machine-like error patterns are evident. These differences contribute to phenomena like algorithm aversion and algorithm appreciation, which have sparked considerable research interest in this area. However, literature remains limited, particularly within specialized domains such as medicine and is often loosely connected. Furthermore, the potential influence of environmental factors, such as time pressure, or task-related knowledge and confidence, have received comparatively little empirical attention. Likewise, user-centric studies involving medical professionals, the primary end users of such systems, remain notably scarce in this context. It also has to be noted that variability in terminology, methodological inconsistencies, and typically small sample sizes limit generalizability and transferability of findings across contexts ~\citep{Patel2015, Fleischmann2014}.

\paragraph{Implications for Our Research} In summary, existing research indicates that human-AI collaboration can give rise to both automation and anchoring biases in clinical decision-making, with automation reliance emerging as factor reflective of both biases. To the best of our knowledge, to date, no empirical research has specifically examined how these biases manifest within the field of pathology. Given the previously outlined limited generalizability of findings from related \ac{hci} domains, the context-dependent interplay of environmental/individual factors and cognitive biases, the distinct nature of healthcare as a decision-making environment, and the central role of visual quantification in pathology, a focused investigation is warranted. This work aims to address that gap.

\section{Methods}
In this work, we quantify automation bias and the anchoring effect during human-AI collaboration in computational pathology and investigate the additional impact of time pressure, professional experience and decision confidence on these phenomena through an online experiment with medical professionals. This section outlines the methodology, data and tools employed in our research.

 \subsection{Study Task}
Our study was conducted in the domain of computational pathology, focusing on the estimation of \acf{tcp} from H\&E-stained tissue slides. \ac{tcp}, typically expressed as a single percentage value or range, refers to the proportion of neoplastic (i.e., tumor) cells relative to the total cellular composition, including stromal, neoplastic and inflammatory cells~\citep{Dufraing2019}. It is important to note that the evaluation is based on the number of neoplastic cells present, rather than the area they occupy~\citep{Dufraing2019}. This distinction presents a challenge, as cell types differ markedly in size, for instance, epithelial cells are substantially larger than inflammatory cells. Furthermore, tumor aggressiveness can further influence epithelial cell size, adding another layer of complexity to accurate estimation.

Although manual cell counting offers greater accuracy, its time-consuming nature makes it impractical for routine use. As a result, brief visual estimation has become the standard, enabling assessments to be conducted even under significant time constraints~\citep{Viray2013}. In our experiment, participants provided single percentage estimates of the \ac{tcp} along a spectrum ranging from 0 to 100 percent. For this purpose, they viewed digitized tissue patches via a web interface equipped with a zoom-function, mirroring standard microscopic evaluation of \ac{tcp} within predefined regions of a tissue slide~\citep{Viray2013}. This setup closely aligns with real-world use cases in pathology archiving and communication systems, where AI support is likely to be deployed~\citep{Zhang2024}.

The primary motivation for centering the study around \ac{tcp} evaluation was the task’s inherent complexity. Although the estimation process itself can be performed fairly quickly, variability in clinical specimens, the lack of standardized protocols, and limited formal training contribute to inconsistent practices across laboratories~\citep{Mikubo2019}. These challenges, coupled with the difficulty of visual quantification for human experts~\citep{Noguerol2019}, an area where AI excels, result in \ac{tcp} estimation being widely recognized for its high interobserver variability across various scientific publications~\citep{Mikubo2019,Smits2014,Bellon2011} and raise concerns about the risk of introducing cognitive biases.

A second rationale for selecting \ac{tcp} estimation as a study task, is its prevalence in routine pathology as well as the adverse clinical implications associated with inaccurate assessment. With the growing prominence of molecular analysis and its establishment as standard practice across many cancer types, precise estimation of tumor cell percentage is essential, as each molecular assay possesses a unique sensitivity threshold requiring a specific level of tumor DNA to ensure robustness and accurate interpretation of test results~\citep{Mikubo2019}. As assay outcomes are crucial for guiding systemic therapy decisions in individual patients, both the testing procedures and the underlying \ac{tcp} estimations must be as precise and reliable as possible, particularly near the assay threshold~\citep{Dufraing2019}. Overestimation of \ac{tcp} in specimens that, in fact, fall below the required threshold may yield false-negative assay results, fostering unwarranted confidence and potentially resulting in misguided treatment efforts. Conversely, underestimation of \ac{tcp} in samples that contain sufficient tumor DNA may also contribute to diagnostic overconfidence, although the downstream clinical implications in such cases are generally less severe~\citep{Smits2014}. Inaccurate \ac{tcp} assessment can also affect numerous other diagnostic disciplines, including the detection of chromosomal translocations, evaluation of copy number variations, determination of residual cancer burden after neoadjuvant therapy, and other genomic aberrations, which might be influenced by contaminated normal DNA~\cite{Viray2013}.

 \subsection{Dataset and AI Algorithm} \label{sec:AI}
For the \ac{tcp} estimation, we provided participants with a selection of 23 tissue patches, presenting a broad spectrum of tumor cellularity and tissue types at different magnifications. Three images were reserved for the training session, while the remaining 20 constituted the main study set. To increase generalizability of the study, data diversity is of high importance. For this reason, we sourced the images from three openly available datasets (all accessible under the Creative Commons license), each featuring high-resolution patches with dense annotations of diverse cell types, including tumor cells: the BreCaHad breast cancer dataset~\citep{BreCaHad}, the colorectal cancer dataset from Frei et al.'s publication on tumor cell fraction scoring \citep{Frei2023}, and the BreastPathQ breast cancer dataset~\citep{Martel2019}.

Owing to the heterogeneous origins of the individual datasets, the image patches vary in size. They were intentionally left untrimmed and unnormalized for presentation to pathologists, in order to incorporate varying difficulty levels through differing image dimensions and mirror challenges typically encountered in real-world \ac{tcp} estimation scenarios. For instance, the selected patches included examples with varying degrees of necrosis and immune infiltrates, differing staining properties, prominent tumor regions intermixed with non-neoplastic cells, and cases with reduced cellularity or low tumor cell fractions. Moreover, the openly available ground truth (GT) annotations in these datasets were reportedly verified by at least one expert pathologist, thereby enhancing their reliability. The complete study dataset, alongside a table with detailed information for each image patch, is available in the supplementary material.

A standard object detection approach based on the FCOS architecture~\citep{Tian2019fcos} was employed to identify and differentiate between tumor and other, non-neoplastic cells, from which the predicted \ac{tcp} for each image patch was calculated. In contrast to anchor-based detectors such as RetinaNet~\citep{Lin2018focal} and Faster R-CNN~\citep{Ren2016faster}, FCOS performs anchor-free, per-pixel predictions of object classification, bounding box regression, and centerness, thereby reducing the complexity associated with anchor design and hyperparameter tuning. A ResNet-18~\citep{He2016ResNet} backbone pretrained on ImageNet~\citep{Deng2009ImageNet} was used as the feature extractor. Feature maps from the C3–C5 stages were utilized to construct a feature pyramid (P3–P7) through 1×1 and 3×3 convolutional projections, each with 256 channels. Both the classification and regression subnets consisted of four stacked 3×3 convolutional layers with 256 filters and ReLU activations, followed by separate prediction heads for class scores, bounding box coordinates, and centerness. The detector operated on 512×512 pixel input patches. The model was trained on the BreCaHad dataset, with five patches from the test split being reserved for inclusion in the experiment. The remaining data were randomly partitioned into training, validation, and test subsets. Original annotations were consolidated: tumor cells retained their labels, while non-neoplastic cells (e.g., apoptotic and tubule cells) were combined into a single non-tumor category. Training was performed using stochastic gradient descent (SGD)~\citep{Loshchilov2017sgdr} with a cosine annealing schedule with warm restarts (CAWR) to adjust the learning rate, which was set to a maximum of 0.001. The model was trained for 200 epochs and repeated five times with different random initializations. The version achieving the highest average precision (AP) on the test set was selected. The use of a fully operative AI system avoided potential bias from researcher-generated "model" predictions and enabled observation of participant behavior in response to realistic automation errors. When applied to the study patches, approximately half yielded highly accurate detections, while the remainder exhibited substantial positive and negative error rates, reflecting realistic non-robustness due to covariate data shifts between datasets.

\subsection{Hypotheses}
Drawing on findings from related literature, the following hypotheses were established:

\subsubsection*{Automation Bias}
\noindent\textbf{H1a:} The introduction of AI support, compared to the baseline (no AI), will lead to the emergence of new errors in the form of measurable automation bias (negative consultations).

\noindent\textbf{H1b:} The presence of time constraints will increase both the frequency and severity of automation bias, as delineated in H1a.

\subsubsection*{The Anchoring Effect}
\noindent\textbf{H2a:} The integration of AI support will produce a measurable anchoring effect, evidenced by a systematic shift from baseline estimates (no AI) toward the AI-generated predictions during AI-aided assessments.

\noindent\textbf{H2b:} Time pressure will amplify the magnitude of the anchoring effect as described in H2a, leading to greater alignment with AI recommendations during AI-assisted evaluations.

\subsubsection*{The Impact of User Characteristics}
\noindent\textbf{H3a:} Automation reliance during \ac{tcp} estimation with AI, representative of both automation and anchoring bias, will be lower among participants with higher levels of self-reported professional experience.

\noindent\textbf{H3b:} Higher levels of decision confidence during the baseline assessment (self-efficacy) will be reflected in reduced reliance on system advice during AI-assisted evaluations.

\noindent\textbf{H3c:} Increased levels of confidence in AI-assisted judgments will be associated with heightened reliance on system advice during AI-supported \ac{tcp} estimation.

\subsection{Experimental Design}
\noindent To investigate the proposed hypotheses, the study followed a $2\times2$ factorial, within-subjects design with two independent variables (IVs): inclusion of AI (yes/no) and presence of time pressure (TP) (yes/no). Automation and anchoring bias constitute the key dependent variables (DVs) in this study. To support their analysis, additional measures were collected, including reliance on AI advice, participants’ professional experience, task performance, and decision confidence.

\paragraph{Automation Bias}\label{sec:abMeasure}
Commencing with the first key measure, automation bias, more specifically its occurrence, was quantified following the methodology proposed by~\cite{Goddard2012}, using the acceptance rate of negative consultations. For this, we identified instances, where an initially correct \ac{tcp} estimate in the baseline condition ($\mathrm{Est}_\textrm{B}$) was altered to a false assessment in the AI-aided evaluation ($\mathrm{Est}_\textrm{AI}$) after exposure to an erroneous AI prediction ($\mathrm{Pred}_\textrm{AI}$). To determine the correctness of \ac{tcp} estimates, we utilized the 25\% threshold required for direct DNA sequencing~\citep{Smits2014}. An assessment was regarded as correct if it matched the GT in relation to this threshold. For the automation bias analysis, the dataset was filtered to only include entries where the independent estimate aligned with the GT in respect to the 25\% marker, while both the AI prediction and the participant’s AI-assisted assessment were incorrect. Following this step, the remaining data tuples were set in relation to the total number of AI-assisted \ac{tcp} evaluations to compute the occurrence frequency of automation bias.

\paragraph{Performance}
Participant performance ($\overline{\left | \mathrm{Dev}_{\textrm{GT}i}\right |}$), defined as the mean absolute deviation of \ac{tcp} estimates from the ground truth (Equation~\ref{eq:perf}), was calculated using the full dataset to evaluate the effect of AI integration on overall accuracy, and separately for instances of negative consultations to assess the severity of automation bias. 

\begin{equation} \overline{\left| \mathrm{Dev}_{\textrm{GT}i}\right|} = \frac{1}{M} \sum_j^M \left |  Est_{ij} - GT_{j} \right | 
\label{eq:perf}
\end{equation}

\noindent Employing the techniques described above, analyses of participant performance, automation bias occurrence and severity were repeated on data subsets stratified by the presence or absence of time constraints to examine the influence of time pressure.

\paragraph{The Anchoring Effect}
The second key dependent variable, the anchoring effect, was defined as a systematic shift from the unaided \ac{tcp} assessment toward the AI-generated prediction in the final, AI-aided judgment. First, this effect was quantified via a descriptive analysis of the mean normalized \ac{woa} metric, as outlined in Section~\ref{sec:bgAE} (Equation~\ref{eq:relAI2}), which reflects the degree of adjustment in response to AI input on a scale from 0 (total disregard) to 1 (full adoption):
\begin{equation}
Rel_\textrm{AI,ij} = \frac{\left | Est_\textrm{AI,ij}  - Est_\textrm{B,ij} \right |}{\left | Est_\textrm{AI,ij}  - Pred_\textrm{AI,ij} \right | + \left | Est_\textrm{AI,ij}  - Est_\textrm{B,ij} \right |}
\label{eq:relAI2}
\end{equation}
It should be noted that, to ensure meaningful computation of this metric and to prevent division by zero, entries in which the independent, AI-generated, and AI-assisted estimates were identical were excluded from the analysis.

Secondly, to test for statistical significance and to ascertain the reliability of the computed $Rel_\textrm{AI,ij}$ measure, which forms the basis for subsequent analyses, we conducted a linear mixed-effects model evaluation. The model was fitted using the lme4 package (V. 1.1.35.5) in R (V. 4.4.0) with p-values for the fixed-effect coefficients being obtained via the lmerTest package (V. 3.1.3) with t-tests based on Satterthwaite’s method. This approach accounted for repeated measures due to the within-subject design of our study and individual differences among participants. Building on prior literature, the LMM approximated AI-aided medical decision-making using the weighted averaging method described in Section~\ref{sec:bgAE} (Equation~\ref{eq:lmmTerm}). Accordingly, the final, AI-supported judgment served as the dependent variable, with the independent estimate and AI prediction included as fixed effects and the participant identifier incorporated as a random effect (random intercept).
\begin{equation}
\mathrm{Est}_{\mathrm{AI}_{ij}} = \beta_0 + \beta_1 \mathrm{Est}_{\mathrm{B}_{ij}} + \beta_2 \mathrm{Pred}_{\mathrm{AI}_{ij}} + u_{0i} + \varepsilon_{ij}
\label{eq:lmmTerm}
\end{equation}
We expect the model to yield a positive coefficient for the AI advice predictor, indicating a shift toward the system prediction in the AI-assisted evaluation. A total of three LMMs were used: Model 1 used the entire dataset to quantify the anchoring effect, while Model 1.1 was fitted on data without time pressure, and Model 1.2 on data with time pressure to examine the influence of time stress.

\paragraph{Professional Experience}
As a supporting measure, professional experience was assessed via a pre-study questionnaire in which participating pathologists self-reported their years of relevant field experience. Response options were grouped into four categories: less than 5 years, 5 to 10 years, 10 to 15 years, and more than 15 years.

\paragraph{Decision Confidence}
Participants rated their confidence on a 5-point Likert scale, ranging from 1 (“not at all confident”) to 5 (“completely confident”), upon submission of their \ac{tcp} assessment for a given tissue patch. Confidence scores were collected in both the baseline and AI condition, with confidence ratings from the independent evaluations acting as an indicator of participants' self-efficacy.

\paragraph{Automation Reliance}
Starting with environmental factors, the influence of time pressure on automation and anchoring bias was conducted using automation reliance as a proxy, given that this metric is indicative of both cognitive biases. The normalized \ac{woa} (Equation~\ref{eq:relAI2}), previously utilized for descriptive evaluation of the anchoring effect, quantifies the degree of assimilation towards external information, consequently also serving as a measure of automation dependence ($Rel_\textrm{AI,ij}$). As before, to circumvent zero division, entries, in which the independent, AI-generated, and AI-assisted estimates are identical, were excluded from the analysis. A two-tailed paired-samples t-test was performed on participant’s mean $Rel_\textrm{AI,ij}$, comparing AI-supported \ac{tcp} assessment with and without the presence of time constraints.

Turning to user characteristics, the relationship between professional experience, decision confidence, and the manifestation of automation and anchoring bias was examined in a secondary linear mixed-effects model evaluation (Model~2), with $Rel_\textrm{AI,ij}$ as the dependent variable (see Equation~\ref{eq:lmmModel2}). Consistent with the previous analysis, participant identifiers were incorporated as a random intercept to account for individual differences and the nature of our study design. Professional experience ($\mathrm{Exp}_{i}$) and decision confidence during both the baseline ($\mathrm{Conf}_{\mathrm{B}_{ij}}$), mirroring self-efficacy, and AI-supported \ac{tcp} assessments ($\mathrm{Conf}_{\mathrm{AI}_{ij}}$) were included as fixed effects. As described earlier, participants self-reported their professional experience by selecting from four ordinal options: 0–5 years, 5–10 years, 10–15 years, and over 15 years. These categories were dummy coded as 0, 1, 2, and 3, respectively, with higher values representing greater expertise. Confidence ratings during both the independent and AI-aided evaluations were originally collected on a Likert scale ranging from 1 to 5. For the LMM investigation, these confidence scores were transformed to a 0–4 scale to align with the coding scheme of professional experience and to establish a meaningful zero point, facilitating interpretability of the model intercept. The magnitude of the fixed effect coefficients indicates the strength of their influence on automation and anchoring bias, while the sign of each coefficient denotes whether the effect is mitigating (negative) or exacerbating (positive).
\begin{equation}
\mathrm{Rel}_{\mathrm{AI}_{ij}} = \beta_0 + \beta_1 \mathrm{Exp}_{i} + \beta_2 \mathrm{Conf}_{\mathrm{B}_{ij}} + \beta_3 \mathrm{Conf}_{\mathrm{AI}_{ij}} + u_{0i} + \varepsilon_{ij}
\label{eq:lmmModel2}
\end{equation}
\begin{figure*}[!t]
    \centering\includegraphics[width=1\linewidth]{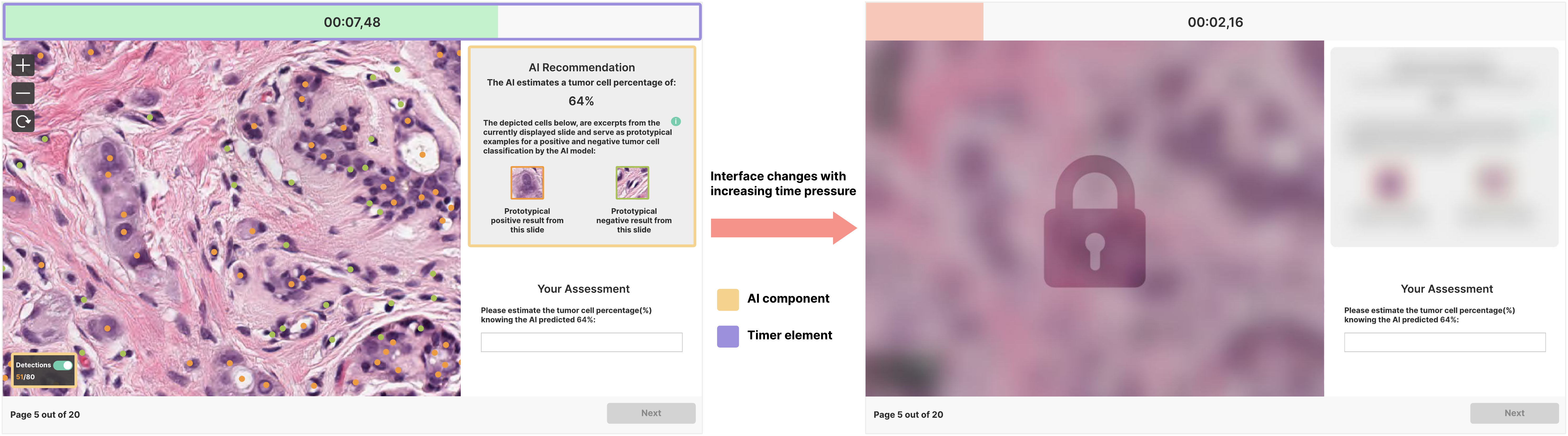}
    \caption{\textbf{
    The study interface presented to participants during \ac{tcp} estimation. Based on the experimental treatment, the AI component (comprising the AI prediction, model reasoning explanations through prototypes, as well as cell detection visualizations) and the countdown timer element were displayed. The arrow indicates how the user interface changed as the timer progressed, generating a sense of urgency and limiting user interaction once the countdown expired.}}
    \label{fig:UI}
\end{figure*}
\subsection{Procedure}
Prior to the main experiment, participants completed a brief training session designed to familiarize them with the user interface and task requirements. Following this, an attention check prompted participants to define \ac{tcp}, ensuring adequate task comprehension and engagement. Afterwards, the main experiment unfolded in two rounds: first pathologists independently estimated the \ac{tcp} across 20 image patches derived from various H\&E-stained tissue slides, then they repeated the task with support from an AI system providing \ac{tcp} predictions. To ensure comparability between the baseline and AI condition, the same image material was used in both experimental phases, which were separated by a two-week washout interval to mitigate learning effects.

Commencing with the second independent variable, half of the slides in each round were evaluated under time constraints, implemented as an expiring 10-second countdown (see Figure~\ref{fig:UI}. Since the time pressure conditions (no time pressure/time pressure) were arranged sequentially within each segment, the study material was divided into two image sets, each encompassing ten distinct yet comparable tissue slides. These sets were curated to match in terms of real-world variability in \ac{tcp} evaluation conditions, such as overstaining, and tumor cell content. Moreover, we balanced representation from the source datasets alongside an equal distribution of AI prediction scenarios, including both under- and overestimations. A within-group design was employed to maximize statistical power, considering the challenges associated with recruiting expert participants, all of whom were required to be actively practicing in the field of pathology. As a result, each subject was exposed to all experimental conditions. To control for order effects, participants were randomly assigned the time pressure condition and image set they begin both rounds with. In addition, the presentation order of tissue patches within each image set was randomized using a balanced 10x10 Latin square.

\subsection{Participants}
Medical experts, including pathologists, pathology residents, and non-physician pathology staff who routinely work with pathology samples in clinical settings, were recruited through an established professional network on a voluntary basis. In accordance with the 2 × 2 factorial study design, the final sample size was constrained to a multiple of four. With the aim to maximize participation, any surplus entries were removed in chronological order. Duplicate entries made by the same subject, as well as participants who failed to submit a complete set of responses for each condition, were excluded from analysis. Out of the original 31 pathology experts who completed the first segment, 28 proceeded to the second round, yielding a final sample size of 28 participants. This final cohort is comprised of 25 pathologists, 2 pathology residents, and 1 non-physician pathology staff member, with the majority reporting over 15 years of professional experience. We independently verified all participating medical experts, confirming their professional roles. Further demographic details are presented in Table~\ref{tab:demograph}. Informed consent was obtained from all participants for the use of their anonymized data in this work.
\begin{table}[h!]
\caption{\textbf{Participant demographics, including age, gender, and professional experience.}}
\label{tab:demograph}
\begin{tabular}{@{} >{\raggedright\arraybackslash}p{0.28\linewidth} 
                >{\raggedright\arraybackslash}p{0.40\linewidth} 
                >{\raggedleft\arraybackslash}p{0.20\linewidth} @{}}
\toprule
\multirow{4}{*}{\textbf{Age}} 
  & 25–34 years & 4 (14\%) \\
  & 35–44 years & 11 (39\%) \\
  & 45–54 years & 8 (29\%) \\
  & \textgreater 55 years  & 5 (18\%) \\
\midrule
\multirow{3}{*}{\textbf{Gender}} 
  & Female              & 9 (32\%) \\
  & Male                & 18 (64\%) \\
  & Prefer not to say   & 1 (4\%) \\
\midrule
\multirow{4}{*}{\textbf{Experience}} 
  & \textless 5 years    & 2 (7\%) \\
  & 5–10 years   & 8 (28.5\%) \\
  & 10–15 years  & 3 (11\%) \\
  & \textgreater 15 years   & 15 (53.5\%) \\
\bottomrule
\end{tabular}
\end{table}
\section{Results}
To ensure the validity of parametric testing, the distributions of pairwise differences between samples used in the t-tests were examined for normality using Shapiro–Wilk tests. In addition, the robustness of the repeated measures ANOVA and linear mixed-effects model analyses was assessed by visually evaluating the normality of residuals for each model. Histograms, scatter plots, and QQ plots indicate that data follows an approximately normal distribution.

\begin{table*}[b]
    \caption{\textbf{Descriptive statistics comparing the occurrence rate of automation bias (accepted negative consultations) and its severity (mean participant performance $\overline{\left | \mathrm{Dev}_{\textrm{GT}i}\right |}$), shown overall and by time pressure condition.}}
    \label{tab:descStats}
    \begin{tabular}{L{0.13\linewidth} P{0.25\linewidth} P{0.27\linewidth} P{0.25\linewidth}}
    \hline
      &  Negative consultations (all) & Negative consultations (no TP) & Negative consultations (TP)\\
    \hline
     Frequency  & 38/560 $\approx$ 7\% & 19/280 $\approx$ 7\% & 19/280 $\approx$ 7\%\\
     \hline
     \rule{0pt}{3ex}$\overline{\left | \mathrm{Dev}_{\textrm{GT}i}\right |} \pm \textrm{SD}$ & 23.61 $\mp$ 19.32 & 19.42 $\mp$ 17.37 & 27.79 $\mp$ 20.71\\
     \hline
    \end{tabular}
    \end{table*}
    
\subsection{The Overall Impact of AI Integration and Time Pressure on Performance}
A two-way repeated measures ANOVA conducted on the mean absolute deviation from the ground truth per participant revealed a statistically significant main effect of AI integration on subject performance, as measured by the $\overline{\left | \mathrm{Dev}_{\textrm{GT}i}\right |}$ metric (F(1, 108) = 8.32, p = \textless .01, \(\alpha\) = 0.05). No statistically significant main effect of time pressure (p = 0.19), nor a statistically significant interaction between AI integration and time stress (p = 0.46) was observed. However, when examining the baseline condition in isolation, a one-tailed paired-samples t-test indicated a statistically significant decline in performance under time constraints (no TP: M = 13.55, SD = 4.81; with TP: M = 15.29, SD = 5.00; t(27) = -1.76, p = 0.045, \(\alpha\) = 0.05). Despite not being statistically significant, the difference in performance between AI-aided conditions reflects a similar tend (no TP: M = 11.74; with TP: M = 12.23). These findings suggest that AI integration generally leads to an increase in performance, as indicated by a reduced $\overline{\left | \mathrm{Dev}_{\textrm{GT}i}\right |}$, while time limitations appear to have the opposite effect.

\subsection{Automation Bias}
Due to the reduced sample size following data filtering as described in Section~\ref{sec:abMeasure}, the dependent variables no longer consistently satisfied normality assumptions. Therefore, the automation bias analysis will be based on descriptive statistics. Out of 560 AI-supported \ac{tcp} assessments, medical experts aligned with AI advice contradictory to their initial estimates, pertaining to the \ac{tcp} being below or above the 25\% threshold, in only 67 instances. Of these, 29 were positive consultations, in which provision of correct AI guidance prompted revision of an initially erroneous assessment. The remaining 38 were negative consultations, where a previously accurate judgment was overturned due to erroneous AI output. Consequently, this corresponds to an observed automation bias rate of approximately 7\% (38/560). As for the impact of time pressure, a review of the descriptive statistics (Table~\ref{tab:descStats}) indicates that the frequency of automation bias remains stable at approximately 7\%, irrespective of time constraints. However, time stress appears to notably amplify the severity of automation bias, as reflected by a decrease in participant performance, evidenced by the elevated $\overline{\left | \mathrm{Dev}_{\textrm{GT}i}\right |}$ (no TP: M = 19.42; with TP: M = 27.79).

\subsection{The Anchoring Effect}
A descriptive analysis of the mean normalized \ac{woa} metric ($Rel_\textrm{AI,ij}$: M = 0.51), calculated across all AI-assisted decisions (excluding cases removed to avoid division by zero), indicates a moderate degree of anchoring on system advice.
\begin{figure}[H]
    \centering\includegraphics[width=0.95\linewidth]{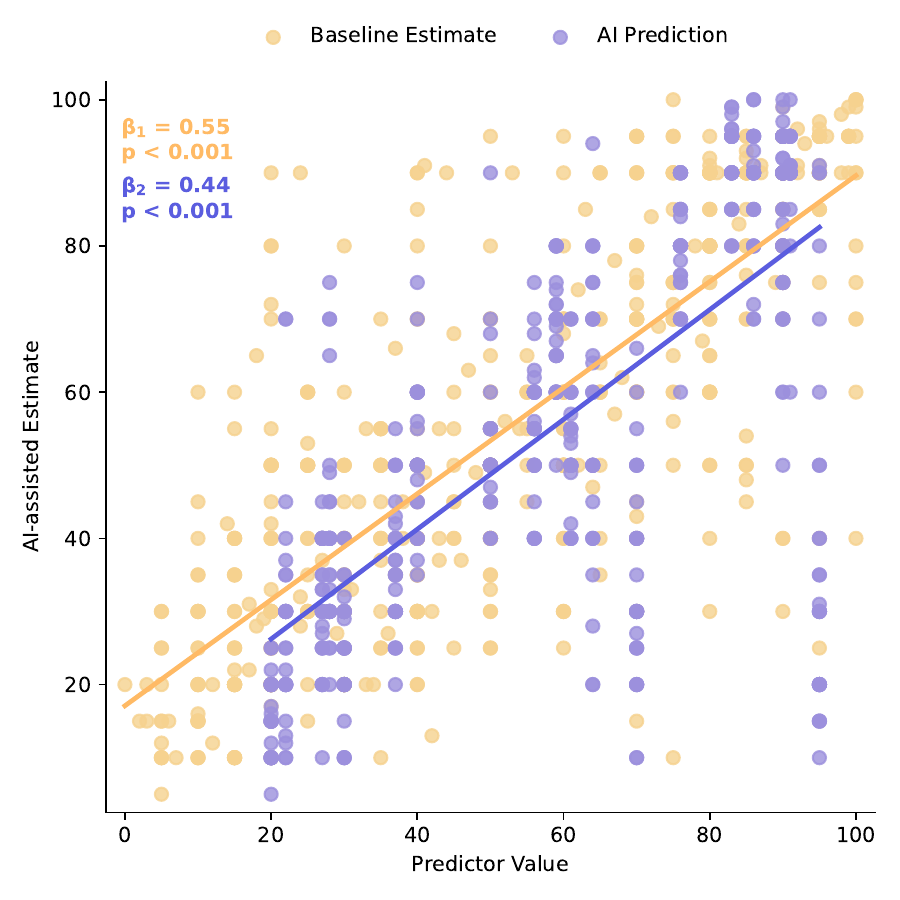}
    \caption{\textbf{Scatter plots illustrating how participants' independent assessments and AI recommendations, respectively, influence the final AI-aided \ac{tcp} estimates. The slopes correspond to the fixed-effect coefficients from the linear mixed-effects model analysis (Model 1), with the sign and magnitude of the coefficient for the system prediction indicating the direction and strength of the anchoring effect.}}
    \label{fig:scatter}
\end{figure}
Turning to parametric testing, Model 1, which utilized the entire dataset (n=560), produced positive and statistically significant regression coefficients for both the independent estimate ($\beta_1$ = 0.55, p~\textless .001) and the AI prediction ($\beta_2$ = 0.44, p~\textless .001) predictor variables (Figure~\ref{fig:scatter}). The coefficient corresponding to the baseline assessment was marginally higher than that of the AI recommendation, indicating a slightly greater weighting of the expert’s initial judgment in the AI-assisted evaluation. However, the coefficients were comparable in magnitude with both being of moderate strength, suggesting that the two factors contributed nearly equally to the formation of final \ac{tcp} estimate. Examination of the coefficient for the AI advice in isolation, reflects the strength and direction of the anchoring effect. The positive value indicates that participants systematically adjusted their assessments toward the AI output, thereby providing quantitative evidence of anchoring on system recommendations.

\begin{figure}[H]
    \centering\includegraphics[width=1\linewidth]{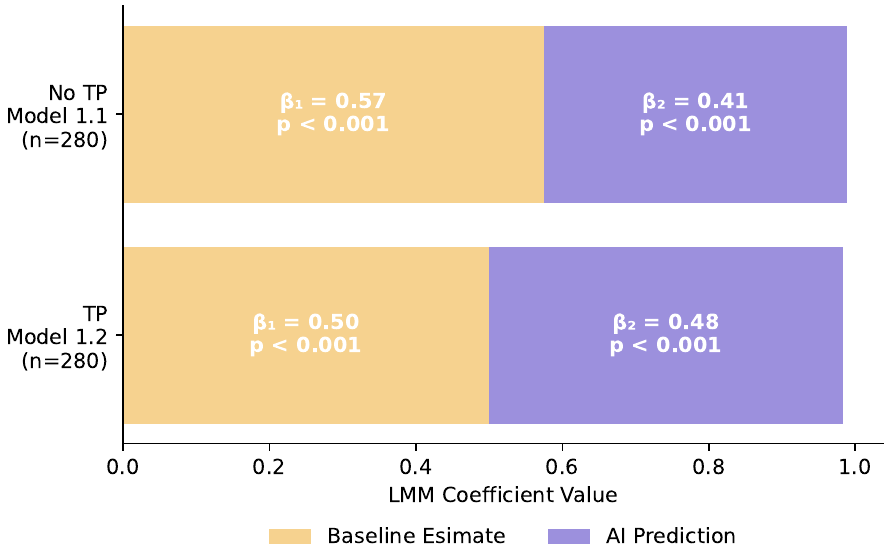}
    \caption{\textbf{Bar plot illustrating differences in linear mixed-effects model coefficients for baseline estimates and AI predictions, comparing AI-assisted assessments conducted with time pressure (TP) (Model 1.2) and without (Model 1.1). The coefficient for AI advice reflects the degree of anchoring on system output.}}
    \label{fig:bar}
\end{figure}

\noindent Taken together, the findings from the descriptive and parametric analyses both indicate a moderate anchoring effect on AI advice, centered around the 0.5 mark. This supports the validity of the $Rel_\textrm{AI,ij}$ metric as a meaningful proxy for anchoring bias/automation dependence, which is critical for its use in subsequent analyses. Importantly, the anchoring effect was observable regardless of whether adopting the AI suggestion would improve or degrade the participant’s independent assessment (for detailed analyses, see Table~\ref{tab:quality} in the appendix).
\begin{table*}[t]
\centering
\caption{\textbf{Results of a linear mixed-effects model analysis (Model) examining how years of professional experience ($\mathrm{Exp}_{i}$), confidence during baseline estimates ($\mathrm{Conf}_{\mathrm{B}_{ij}}$), reflecting self-efficacy, and confidence during AI-aided evaluations ($\mathrm{Conf}_{\mathrm{AI}_{ij}}$) respectively affect automation reliance ($\mathrm{Rel}_\textrm{AI,ij}$), a measure indicative of both automation and anchoring bias. Professional experience, self-reported by participants using four ordinal categories (\textless 5, 5–10, 10–15, and \textgreater 15 years), was dummy coded from 0 to 3, with higher values denoting greater experience. Confidence ratings, originally measured on a five-point Likert scale (1: not at all confident - 5: completely confident), were rescaled to a 0–4 range to align with the experience coding and to enable meaningful interpretation of the model intercept.}}
\label{tab:traits}
\begin{tabular}{L{0.15\linewidth}P{0.11\linewidth}P{0.11\linewidth}P{0.11\linewidth}P{0.11\linewidth}P{0.11\linewidth}P{0.11\linewidth}}
\hline
\multicolumn{7}{c}{Model 2 (n=549)}\\
\hline
\textbf{Variable} & \textbf{Coef. ($\beta$)} & \textbf{SE} & \textbf{t} & \textbf{p}  & \multicolumn{2}{c}{\textbf{95\% CI}}\\
 & & & & & \textbf{LL} & \textbf{UL} \\
\hline
Intercept & 0.51 & 0.07 & 7.43 & \textless 0.001 & 0.38 & 0.64 \\
$\mathrm{Exp}_{i}$ & -0.06 & 0.02 & -3.30 & \textless 0.001 & -0.09 & -0.02\\
$\mathrm{Conf}_{\mathrm{B}_{ij}}$ & -0.03 & 0.02 & -1.97 & 0.050 & -0.07 & 0.00 \\
$\mathrm{Conf}_{\mathrm{AI}_{ij}}$ & 0.08 & 0.02  & 3.71  & \textless 0.001 & 0.04 & 0.12 \\
\hline
\end{tabular}
\end{table*}
To examine the impact of time stress, the dataset from Model 1 was split based on the presence or absence of time constraints during \ac{tcp} assessment, as outlined in the experimental design. The LMM analysis of anchoring bias was then repeated on each subset, resulting in Models 1.1 (no time pressure) and 1.2 (with time pressure). Comparison of the coefficients from Model 1.1 ($\beta_1$ = 0.57, p~\textless .001; $\beta_2$ = 0.41, p~\textless .001) and Model 1.2 ($\beta_1$ = 0.50, p~\textless .001; $\beta_2$ = 0.48, p~\textless .001) depicted in Figure~\ref{fig:bar}, all positive and statistically significant, indicates that the independent estimate remains the slightly stronger influence on AI-assisted decision-making across time pressure conditions. However, time strain appears to increase the weight of AI advice, as reflected by the higher coefficient for the AI prediction in Model 1.2 compared to Model 1.1. This suggests that anchoring effect becomes more pronounced under time stress.

\subsection{The Influence of Time Pressure and User Characteristics on Automation Reliance}
While the preceding LMM analysis already indicated increased automation reliance under time constraints, this observation was further examined through a two-tailed paired-samples t-test on the mean normalized \ac{woa} metric ($Rel_\textrm{AI,ij}$) per participant (excluding cases removed to avoid division by zero). The t-test revealed a statistically significant difference between the AI treatments with and without time pressure (no TP: M = 0.48, SD = 0.13; with TP: M = 0.54, SD = 0.13; t(27) = 2.55, p = 0.017, \(\alpha\) = 0.05). This finding highlights that dependence on AI advice may intensify under time stress.

\paragraph{Impact of Person-specific Traits}
Shifting focus to user characteristics, although confidence ratings were collected using an ordinal Likert scale, ~\cite{Norman2010} has demonstrated that parametric analyses are appropriate in practice when the data is normally distributed and each group includes more than 5–10 observations. Accordingly, a one-tailed paired-samples t-test was performed on participants’ average confidence scores, comparing the baseline and AI-assisted conditions. The results suggest that AI integration may generally increase participants’ confidence in their \ac{tcp} assessments  (Baseline: M = 3.30, SD = 0.65; with AI: M = 3.64, SD = 0.58; t(27) = 3.49, p =~\textless .01, \(\alpha\) = 0.05).

On the flip side, it is also conceivable that the higher decision confidence during AI-assisted assessments could lead to greater susceptibility to anchoring. To explore this hypothesis, along with the potential influence of professional experience and self-efficacy, a second linear mixed-effects model analysis (Model 2) was conducted. Time pressure was not incorporated as a predictor into this LMM, as it had already been examined separately due to its categorical format. The model’s intercept ($\beta_0$ = 0.51, p~\textless .001) represents the expected level of automation reliance when all predictor variables are held at zero, providing a reference point for interpreting the effects of predictors, all exhibiting statistically significant coefficients (see Table~\ref{tab:traits}). The negative coefficient associated with professional experience ($\beta_1$ = -0.06, p~\textless .001) indicates that an increase in  clinical experience weakens the dependence on AI recommendations. A similar pattern emerged for baseline decision confidence (self-efficacy: $\beta_2$ = -0.03, p= 0.050), with higher confidence levels being linked to reduced reliance on AI advice. In contrast, decision confidence during AI-assisted evaluations yielded a positive coefficient ($\beta_3$ = 0.08, p~\textless .001), suggesting that heightened confidence in the presence of AI support may actually intensify automation dependence.

\section{Discussion}
Although the impact of AI recommendations on medical decision-making is gaining research attention, user-centered studies that quantify these effects are still scarce. This study seeks to address an existing gap by empirically investigating the extent to which AI integration may induce automation and anchoring biases, while additionally assessing how time pressure, professional experience, and decision confidence shape this interaction.

In general, AI integration was found to enhance participant performance. This beneficial effect may also help explain why the pronounced performance decline observed under time pressure in the baseline condition was less evident during AI-assisted evaluations. However, despite the potential advantages of AI support, pathologists seemed reluctant to accept model suggestions that contradicted their initial assessments, regardless of the correctness of the system output. Within the scope of this study, automation bias does not appear to be the predominant cognitive bias influencing AI-assisted medical decision-making. Nevertheless, by quantifying accepted negative consultations, we identified an automation bias occurrence frequency of approximately 7\% in its purest form. This is consistent with previous empirical work reporting acceptance rates of negative consultations ranging from 6\% to 11\%~\citep{Goddard2012}. Since our results show that AI interaction can indeed give rise to automation bias, leading to errors unlikely to occur without system input, Hypothesis H1a is fully accepted. Contrary to expectations outlined in H1b, the presence of time stress did not impact the frequency of automation bias occurrence. This outcome might stem from the relatively modest number of automation bias cases in our sample, which may have been insufficient to reliably detect such effects. It may also relate to the nature of our time pressure simulation. In clinical practice, time constraints typically arise from high case volumes rather than fixed countdowns, which may limit the extent to which participants' behavior in our study mirrors real-world responses under time strain. However, time pressure appears to amplify the severity of automation bias, as reflected in the pronounced decline in average participant performance under time stress, beyond what would be expected from the negative impact of time constraints on performance alone (as shown in the general analysis). This notion may be explained by the t-test results, which revealed a statistically significant increase in automation reliance under time pressure, evidenced by a higher mean normalized \ac{woa} value. Consequently, the heightened alignment with erroneous AI output under time strain may exacerbate the severity of automation bias. In summary, although we did not observe a significant effect of time constraints on the frequency of automation bias, our findings suggest that its severity worsens when time is limited. Therefore, Hypothesis H1b is partially accepted.

Turning to our second set of hypotheses, both the descriptive analysis and the linear mixed-effects model results indicate an anchoring effect of moderate strength, centering around the 0.5 mark. Furthermore, the positive and statistically significant LMM coefficient for the AI prediction suggests that participants systematically adjusted their AI-aided \ac{tcp} assessments toward the system output, weighting it nearly as strongly as their own independent estimates, which however remained the primary decisive factor.~\cite{Bailey2023} suggest that rational advice use would involve adjusting one’s judgment about 50\% toward external input. In practice, however, adoption rates tend to hover around 30\%, a phenomenon known as advice discounting. Our observed anchoring effect exceeds this benchmark. Similarly,~\cite{Logg2019} reported a mean (windsorized) \ac{woa} of approximately 0.45 in the context of algorithm-assisted judgments. This suggests that while individuals still weigh their own assessments more heavily than third-party input, algorithmic advice may be adopted more readily than human recommendations in traditional advice-taking paradigms. This phenomenon, known as algorithm appreciation, is driven by the perception of algorithms as competent, objective, and consistent, building trust and potentially fostering overreliance on automation. While this could even be beneficial when the system is highly accurate, it may be detrimental to patient safety when AI errs. Consistent with this concern, supplementary analyses presented in the appendix revealed a statistically significant anchoring effect on AI predictions regardless of system accuracy. Although the effect was attenuated when AI advice was of lower quality, it remained reliably present, suggesting that participants were not fully able to disregard inaccurate AI suggestions, even when they likely recognized their diminished reliability. In summary, Hypothesis H2a, stating that the presence of AI produces a measurable anchoring effect, reflected as a systematic shift toward system predictions during human AI collaboration, is accepted. As for the influence of time stress, while the regression coefficients for Models 1.1 and 1.2 are all positive, statistically significant and of moderate strength, the independent \ac{tcp} estimates remain the primary factor driving AI-assisted evaluations regardless of time constraints. However, the increase in the AI prediction coefficient from Model 1.1 to Model 1.2 suggests that time pressure amplifies the influence of system advice on final judgments, resulting in stronger anchoring on AI recommendations. These results mirror the aforementioned t-test findings, which indicate increased automation reliance under time limitations. This observation may be explained by the idea that time stress strains cognitive resources, prompting individuals to lean more heavily on system suggestions to reduce mental load. Based on these insights, we fully accept Hypothesis H2b, which posits that time pressure exacerbates the anchoring bias.

Results of the secondary LMM analysis indicate that higher levels of self-reported professional experience are associated with reduced automation reliance, a proxy we used for both automation and anchoring bias. This is reflected in a statistically significant negative regression coefficient, suggesting that more experienced practitioners might tend to place greater trust in their own clinical judgment. Their extensive domain knowledge may foster a more critical stance toward external input, including AI recommendations.

Decision confidence during the baseline assessments, a reflection of self-efficacy, also yielded a statistically significant, negative model coefficient, suggesting that dependence on system output weakens with greater self-efficacy. This highlights that medical professionals with higher initial confidence in their independent judgments may feel less inclined to defer to AI guidance. Conversely, automation reliance seems to grow in tandem with decision confidence during AI-assisted evaluations, as evidenced by a statistically significant, positive regression coefficient. These observations indicate that when users feel reassured by the presence of AI support, they may increasingly align their decisions with the system’s output. Our findings on the influence of user characteristics on the manifestation of automation and anchoring bis are in line with prior research. Accordingly, Hypotheses H3a and H3b, which propose that professional expertise and self-efficacy respectively can buffer against automation reliance, and Hypothesis H3c, which posits that greater confidence during AI-supported medical decisions may increase system dependence, are supported.

To summarize, this study demonstrated the presence of automation and anchoring bias in AI-assisted medical decision-making in pathology, while also examining how environmental factors and user characteristics modulate these biases. Automation reliance emerged as a key indicator of both effects. However, it is important to recognize that reliance on automation is not inherently undesirable. Despite its link to cognitive biases, it can also support diagnostic accuracy when AI predictions are of high quality. For example, although greater professional experience was associated with reduced automation reliance, and thus a lower susceptibility to AI-induced bias, it may also lead to underutilization of accurate system advice. These findings underscore the importance of fostering appropriate reliance through e.g., user training and system design, encouraging users to trust and adopt accurate AI guidance while remaining vigilant and discerning when system outputs are flawed. Establishing such calibrated trust is essential to realizing the full benefits of human-AI collaboration in healthcare.

\subsection{Limitations and Future Work}
We acknowledge several limitations of this study. First, the limited availability of medical experts resulted in a relatively small sample size (n = 28). Although a within-subject design was employed to enhance statistical power, the observed effects may still be under- or overrepresented due to sample-specific variability, limiting the generalizability of the findings to the broader target population. Additionally, the recruitment of participants from an existing professional network may further constrain the representativeness of the sample. Nonetheless, it is worth noting that, despite its modest size, the sample is consistent with those reported in related studies within the \ac{hci} domain~\citep{Calisto2023, Bach2023, Bascom2024}. Second, user interface design choices, such as the omission of clinical background information and the lack of alternative slide views, may have decreased task realism. As a result, participant behavior may have been altered, prompting subjects to complete the study with less diligence than their routine diagnostic work, affecting the manifestation of observable automation and anchoring bias. Additionally, the inclusion of prototypes, an explainable AI (xAI) technique, as part of the user interface may have influenced participant responses, as system explanations are known to foster trust~\citep{Schemmer2022}. This increased trust could have contributed to a more positive mental model of the AI system, potentially elevating automation reliance and, consequently, the occurrence of automation and anchoring bias. Thirdly, as described in Subsection \ref{sec:AI}, the AI system produced highly accurate predictions on roughly half of the study slides, while the remaining cases showed varying degrees of error. Participants likewise varied in performance, with some consistently outperforming the AI and others falling below its accuracy, reflecting real-world variability in both human expertise and AI robustness. Our supplementary analysis (Table \ref{tab:quality}) shows that anchoring persists even when the AI system provides low-quality advice (i.e., predictions that deviated more from the ground truth than participants’ independent estimates), albeit being less pronounced. Moreover, we observed a 7\% occurrence rate of automation bias in the form of accepted negative consultations, aligning with the 6–11\% range reported across other medical specialties in Goddard et al.’s review article~\citep{Goddard2012}. This convergence suggests that automation bias displays a notable degree of robustness across clinical contexts and study configurations. Taken together, our findings indicate that automation and anchoring bias are likely to emerge across a variety of human–AI performance constellations. However, it is important to acknowledge that the magnitude of these effects depends, among other factors, on the relative accuracy of the clinician and the AI system in question. Consequently, our outcomes are partly shaped by the specific conditions of our study, including the individual capabilities of the medical expert and algorithm, as well as the definition of a “beneficial” AI consultation for analytical purposes, and may manifest differently in other study designs or real-world decision-making contexts. Lastly, the aforementioned challenge of simulating time pressure may not have accurately captured participants’ reactions and their reliance on AI support under real-world clinical time constraints. In our study, time stress was induced using an expiring timer for each tissue patch, chosen for the controllability it provided in a remote setting. However, this method differs from the typical experience of time strain in routine pathology, where deadlines are shaped by the volume of specimens to be assessed rather than individual countdowns. While this study focused primarily on temporal demands, it should be noted that daily clinical stress is influenced by a broader range of contextual factors, including sleep quality, caffeine intake, emotional state, physical condition, and interpersonal or workplace dynamics, among others. Furthermore, \ac{tcp} estimation, though clinically relevant, does not carry the immediate life-or-death implications present in other medical specialties. As such, our experimental setup captures only the time-based aspect of clinical stress, rather than its full scope as experienced in real diagnostic practice. Despite these limitations, this study offers an important initial step toward a more comprehensive understanding of automation bias and the anchoring effect in AI-assisted medical decision-making, as well as the factors that shape them. As an initial exploration of automation and anchoring bias in the context of \ac{dss} use within computational pathology, this study provides a foundation for future research to examine these cognitive biases in more diverse, realistic, and high-stakes clinical settings, where diagnostic decisions directly impact patient outcomes and cognitive load emerges through multiple channels over extended time periods, given that generalizability is often achieved through collective research efforts. Moreover, future work could also investigate the effectiveness of mitigation strategies, such as cognitive forcing functions, in reducing automation and anchoring bias, particularly under time pressure and in relation to user-specific characteristics.

\section{Conclusion}
In this paper, we empirically examined the influence of AI integration, time pressure, and user-specific traits on the emergence of automation and anchoring bias during interaction with \acp{cdss} in computational pathology. Through a web-based experiment involving 28 trained pathology experts, who estimated tumor cell percentages with and without AI support, partially under time constraints, we found that while AI generally improved diagnostic accuracy, it also introduced measurable cognitive biases. Specifically, approximately 7\% of all decisions made with AI support were subject to automation bias, where participants overturned previously correct assessments in favor of incorrect AI suggestions. Although time pressure did not notably increase the frequency of such occurrences, it appeared to exacerbate their severity, as evidenced by performance degradation linked to increased automation reliance under cognitive load. A linear mixed-effects model analysis further revealed a statistically significant regression coefficient for AI recommendations, indicating a moderate shift, i.e., anchoring, toward system output, an effect that appeared to strengthen under time strain. A secondary LMM evaluation demonstrated that professional experience and self-efficacy mitigated AI dependence, a factor indicative of both automation and anchoring bias, whereas heightened confidence during AI-supported decision-making amplified it. By raising awareness of the potential risks of inducing cognitive biases and examining the factors that shape them, we aim to contribute to laying the groundwork for the safe integration of AI in high-stakes fields such as medicine.


\acks{M.A. acknowledges funding from the German Research Foundation  (DFG, Deutsche Forschungsgemeinschaft, project number: 520330054). M.A. and J.A. acknowledge support by the Bavarian State Ministry of the Sciences and the Arts (project FOKUS-TML). M.A., J.A. and E.R. acknowledge funding by the Bavarian Institute for Digital Transformation (bidt) under the grant ``Responsibility Gaps in Human Machine Interaction (ReGInA)``.

We also acknowledge the use of artificial intelligence tools to improve the writing style and grammar of the manuscript. These edits were made with careful consideration of the original content, aiming to improve the overall clarity and quality. Additionally, the user interface of the study was created using resources from Flaticon.com.}

%
\ethics{This study did not require formal ethics board approval, as it involved a minimal-risk, web-based experiment conducted with expert participants. The task, estimating tumor cell percentages, is a routine part of these professionals' daily practice. No sensitive data, invasive procedures, or vulnerable populations were involved. All participants were fully informed about the study’s nature and structure, and their participation was entirely voluntary. All materials and data used in this research were sourced from publicly available datasets, ensuring transparency and adherence to ethical research standards.}

\coi{The authors declare no conflicts of interest related to this work.}

\data{The source code for the experimental platform is available on GitHub: \url{https://github.com/emelyrosbach/SoS}. The study image material, including a table with detailed image patch information, recorded participant data, and the code used for statistical analyses, are provided in the supplementary material of this paper. All materials are made available under the CC BY 4.0 license.}

\bibliography{sample}


\appendix
\onecolumn 
\section{The Influence of AI Prediction Quality on Anchoring Behavior}
    \begin{table} [H]
    \caption{\textbf{Linear mixed-effects model results comparing the coefficients for baseline estimates and AI predictions in AI-assisted assessments, distinguishing between high-quality AI guidance, where the system’s recommendation was closer to the ground truth than the participant’s independent estimate (Model 3), and low-quality outputs, where the AI suggestion deviated more from the ground truth than the independent assessment (Model 4). The coefficient for AI predictions reflects the degree of anchoring on the system output.\newline}}
    \label{tab:quality}
    \begin{tabular}{L{0.07\linewidth}  P{0.07\linewidth}P{0.05\linewidth}P{0.05\linewidth}P{0.05\linewidth}P{0.05\linewidth}P{0.05\linewidth} |P{0.07\linewidth}P{0.05\linewidth}P{0.05\linewidth}P{0.05\linewidth}P{0.05\linewidth}P{0.05\linewidth}}
    \hline
     & \multicolumn{6}{c|}{\textbf{Model 3: High Quality Advice (n=297)}} & \multicolumn{6}{c}{\textbf{Model 3: Low Quality Advice (n=234)}} \\
    Variable & Coef. & SE & t & p & \multicolumn{2}{c|}{95\% CI} & Coef. & SE & t & p & \multicolumn{2}{c}{95\% CI}\\
     & & & & & LL & UL & & & & & LL & UL\\
    \hline
    Intercept & 0.76 & 1.47 & 0.518 & 0.61 & -2.11 & 3.63 & 2.23 & 2.62 & 0.85 & 0.396 & -2.91 & 7.37\\
    $\mathrm{Est}_{\mathrm{B}_{ij}}$ & 0.26 & 0.03 & 10.10 & \textless 0.001 & 0.21 & 0.31 & 0.64 & 0.03 & 18.41 & \textless 0.001 & 0.57 & 0.71\\
    $\mathrm{Pred}_{\mathrm{AI}_{ij}}$ & 0.74 & 0.03 & 25.36 & \textless 0.001 & 0.69 & 0.80 & 0.30 & 0.04 & 7.74 & \textless 0.001 & 0.22 & 0.37 \\
    \hline
    \end{tabular}
    \end{table}
\end{document}